\documentclass[
]{ceurart}

\usepackage{listings}
\usepackage[english]{babel}

\usepackage[letterpaper,top=2cm,bottom=2cm,left=3cm,right=3cm,marginparwidth=1.75cm]{geometry}

\usepackage{amsmath}
\usepackage{graphicx}

\begin{document}

\copyrightyear{2023}
\copyrightclause{Copyright for this paper by its authors.
  Use permitted under Creative Commons License Attribution 4.0
  International (CC BY 4.0).}

\title{Recording of 50 Business Assignments}
\conference{}
\author[1]{Michal Sroka}[%
orcid = 0000-0002-7505-2521,
email=misroka@microsoft.com
]
\cormark[1]
\author[1]{Mohammadreza {Fani Sani}}[%
orcid=0000-0003-3152-2103,
email=mfanisani@microsoft.com
]
\address[1]{Microsoft Development Center Copenhagen, Denmark}

\begin{abstract}
One of the main use cases of process mining is to discover and analyze how users follow business assignments, providing valuable insights into process efficiency and optimization. In this paper, we present a comprehensive dataset consisting of 50 real business processes. The dataset holds significant potential for research in various applications, including task mining and process automation which is a valuable resource for researchers and practitioners. 
\end{abstract}


\maketitle

\section{Introduction}

Process mining is a field that aims to bridge the gap between classical data science and business process management. One of the main goals of process mining is to improve the process that can be achieved in various ways, e.g., redesigning the process and process automation~\cite{PM4PY}. 

Currently, process automation is an emerging field within process mining that strives to enhance operational efficiency by automating manual tasks. By automating these tasks, organizations can streamline their processes, reduce errors, and increase productivity~\cite{rpa_taskmining_gap}. However, to achieve successful process automation, it is essential to have a comprehensive understanding of the underlying low-level tasks and their interconnections through various application programming interfaces (APIs). These APIs facilitate communication and data exchange between different systems, enabling seamless integration and coordination of automated tasks. By gaining insights into the intricacies of these low-level tasks and their API interactions, organizations can effectively design and implement robust process automation solutions that optimize resource utilization and drive overall process improvement.

Microsoft's Power Connectors \cite{microsoft_power_connectors} act as proxies or wrappers for APIs, enabling seamless communication between Microsoft Power Automate, Microsoft Power Apps, Azure Logic Apps, and underlying services. These connectors allow users to connect their accounts and utilize prebuilt actions and triggers to build applications and workflows. With a wide range of software as a service (SaaS) connectors, Microsoft offers a vast ecosystem that facilitates connecting cloud-based apps, data, and devices. Popular connectors include Salesforce, Office 365, Twitter, Dropbox, and various Google services. By leveraging these connectors, organizations can integrate external services into their business processes, retrieve data, perform actions, and synchronize information effortlessly.

These connectors play a vital role in business process automation by simplifying the utilization of APIs. Acting as intermediaries, they abstract the complexities of direct API integration, providing a user-friendly interface for interacting with external services. The connectors handle authentication and communication protocols, establishing secure and reliable connections with target APIs. This eliminates the need for extensive API expertise, allowing businesses to focus on app and workflow development. Microsoft's Power Connectors empower organizations to streamline processes, enhance productivity, and foster innovation by seamlessly integrating APIs and external services into their operations.

In this paper, we aim to provide some information about the dataset that we gather from recordings of 50 different processes. This dataset can be accessed via github repository \cite{microsoft_50businessassignmentslog}.

In the following, we first explain how we gathered the recordings. Thereafter, some characteristics of this dataset will be discussed and finally we provide a conclusion and some possible use cases of this dataset. 

\section{Dataset}
In this section, we provide an overview of the dataset, including the process of creating it and its key characteristics. 
\subsection{How the dataset was created}
The aim of this dataset is to cover a broad range of business assignments common for office workers across many industries. Such coverage can help in developing new algorithms for Robotic Process Automation (RPA) and task mining~\cite{rpa_taskmining_gap}. Many of the common tasks can be automated using Power Connectors. There are hundreds of Microsoft Power Platform Connectors which can be used for this purpose. The connectors selected for this dataset are among the most commonly used.

Each of the processes aims at presenting a process whose part could be automated using one or more of such connectors. The connectors were selected apriori to the recording. 

The method for creation of the recording is: 
\begin{enumerate}
    \item Select two or more connectors
    \item Select a business scenario which utilized the selected connectors
    \item Create a task description
    \item Perform the task while recording (i.e., recording all the steps that are by the user).
\end{enumerate}

To explain the record creation process, please consider the following example. 
\begin{enumerate}
    \item Considering Sharepoint and Onenote as the selected connectors
    \item Defining a business scenario to use the selected connectors "during the field survey, you collected feedback points. To discuss with team members, you are going to share those points on collaborative platform."	

    \item Describing the task for the user to run the defined business scenario:
    \begin{enumerate}
        \item Go to the personal note taking app
        \item Locate the desire file and copy the points
        \item Go to Sharepoint
        \item Create a document to discuss with Team Members
    \end{enumerate}
    \item Recording the user when they perform each small steps, e.g., clicking or typing. For see the details steps of this example, please refer to $Recording\_91$ of $Process\_33$ in the github repository \cite{microsoft_50businessassignmentslog}.
\end{enumerate}

After all recordings were created, a sequential processID was assigned to each specific process.

\subsection{Dataset Characteristics}
Here, we provide Characteristics information of the gathered dataset. 
The dataset contains $165$ cases (or recordings) that are related to $50$ unique processes. There are $5718$ events (or steps). Each process has at least $3$ cases. In each case, in average there  $35$ events. 
For each event/step available information is summarized in Table 1. 

\begin{table}[tb]
\caption{ Different columns that discribes an event in dataset.}
\begin{tabular}{|l|l|}
\hline
Column Name & Details \\ \hline
StepId & Description: The number of the step within the recording. \\ & Example: 1 \\ \hline
RecordingId & Description: ID of the recording. \\ & Example:  Recording\_4 \\ \hline

ProcessId & Description: ID of the process, for which the recording was done. \\ & Example: Process\_3 \\ \hline

TimeStamp & Description: Timestamp of the step within the recoridng. \\ & Example: 2022-02-21T08:19:12+00:00 \\ \hline

StepName & Description: Name of the step. It is one of 19 different options. \\ & Example: Press button in window \\ \hline

StepDescription & Description: More detailed description of the step. \\ & Example: Button 'New Tab' in Window 'Process ... - Microsoft Edge'\\ \hline

ApplicationProcessName & Description: Process Name, taken from the opened application. \\ & Example: msedge \\ \hline

ApplicationParentWindowName & Description: Parent Window Name, taken from the opened application. \\ & Example: Process advisor | Power Automate and 1 more page \\ & - Personal - Microsoft Edge \\ \hline

AutomationCode & Description: A Script which could be used to automate that step. \\ & Example: [``UIAutomation.PressButton Button: 
\\ &  appmask[‘Window ’Process … - Microsoft Edge’‘]
\\ & [‘Button ’New Tab’’] n’'] \\ \hline

label\_EventName & Description: Event name given by a person making the recording for the\\ & groupped steps.  \\ & Example: Check weather condition. \\ \hline

label\_EventId & Description: ID of the group of steps, called an Event. \\ & Example: 2\\ \hline
\end{tabular}
\end{table}

The activities are accurately labeled by human judges who used different terminology for similar activities, there are $752$ unique activities in this dataset.

\subsubsection{ Data model and schema of the resource }
Here, we explain each column of the dataset. 
\begin{description}
    \item[ProcessId:]  ID of the process. The process is defined on a high level by up to 3 tasks which needs to be achieved. This tasks use various business applications, detailed in the ApplicationParentWindowName column described below.
    
    \item[RecordingId:] ID of the recording within the process.  Every process is typically recorded 3 times. In this context Recording ID could be considered as a Case ID.

    \item[StepId:] ID of the step within the recording of the process. A step refers to a discrete action or operation performed as part of a broader task. It represents a specific unit of work that contributes to achieving the overall goal, and is the most fine grained operation considered such as a click of a mouse or stroke of a key. 
    
    \item[StepName:] Step Name is a standardized name of the action taken by the user. It is one of the 19 values, with the distribution shown in Table \ref{tab:StepName}.

    \item[TimeStamp:] Time when the specific step was recorded, automatically captured by the system. The granularity of the timestamp is in seconds.

    \item[StepDescription:] Extended description of each step automatically generated by the software. E.g. 
    \emph{Check Box 'All day event' in Window 'Untitled - Appointment  '}
    
    \item[ApplicationProcessName:] Name of the process associated with the application. STatistical analysis of this column can be found in Table \ref{tab:ApplicationProcessName}.
    
    \item[ApplicationParentWindowName:]  The name of the Window as shown by the parent application. For example opening accuweather in Microsoft Edge browser, while haveing 2 other tabs opened, is shown as 
    \emph{accuweather - Search and 2 more pages - Personal - Microsoft Edge}

    \item[AutomationCode:] Code which can be used for automating the step.

    \item[NextStepId:] ID of the next step. This Id corresponds to StepID field above. It can be used for chaining steps in the event that a step has been deleted.
    
    \item[labe\_EventName:]  Human label for EventName, which groups certain tasks together into an event.
    
    \item[label\_EventId:] Automatically assigned ID for the Event Name given by the human judge.

\end{description}

\begin{table}[h]
\caption{Step names and their frequency}
\label{tab:StepName}
\centering
\begin{tabular}{|l|l|}
\hline
StepName & Count \\
\hline
Click UI element in window & 2504 \\
Press button in window & 1406 \\
Populate text field in window & 718 \\
Select menu option in window & 404 \\
Send keys & 387 \\
Drag and drop UI element in window & 110 \\
Select tab in window & 75 \\
Set checkbox state in window & 44 \\
Set drop-down list value in window & 20 \\
Select radio button in window & 16 \\
MouseAndKeyboard.SendKeys.FocusAndSendKeys & 15 \\
Expand/collapse tree node in window & 9 \\
Comment & 2 \\
Move window & 2 \\
Prepare a form for employees feedback & 0 \\
Close window & 1 \\
Resize window & 1 \\
Locate the Notification and review it in Inbox of mailing app & 1 \\
Get details of a UI element in window & 1 \\
\hline
\end{tabular}
\end{table}

\begin{table}[h]
\caption{Frequencies of ApplicationProcessName}
\label{tab:ApplicationProcessName}
\centering
\begin{tabular}{|l|l|}
\hline
ApplicationProcessName & Count of steps \\
\hline
chrome & 1661 \\
msedge & 1128 \\
firefox & 884 \\
Teams & 570 \\
PBIDesktop & 125 \\
OUTLOOK & 119 \\
ApplicationFrameHost & 107 \\
Ssms & 46 \\
CoollePDFConverter & 42 \\
SearchApp & 38 \\
OneDrive & 27 \\
ONENOTE & 26 \\
explorer & 21 \\
Skype & 19 \\
EXCEL & 16 \\
ShellExperienceHost & 13 \\
cmd & 3 \\
\hline
\end{tabular}

\end{table}

\section{Conclusion and Use cases}
In this paper, we explain a dataset that aims to provide a comprehensive coverage of business assignments commonly encountered by office workers across various industries. The dataset is can be used to facilitate the development of new algorithms for Robotic Process Automation (RPA), with a focus on utilizing Microsoft Power Platform Connectors. We describe the process of creating the dataset, charecterization information of it, and explanation of its main fields. Overall, this dataset serves as a valuable resource for advancing research in the field of task mining, and automation. 

The current event log can be used for various use cases. In the following, we explain two possible use cases.
\subsubsection*{Process Mining for Automation}
One of the important scenarios that process mining can help industries to reduce the required times and cost is to automate their processes. In this regard, using process mining we are able to detect the common frequent patterns in the processes and recommend them for possible automation. 

As explained in this dataset, we have at least $3$ recordings that are done by different humans for doing similar tasks. Using this dataset, we are able to assess how different automation detection methods are able to detect frequent patterns and also what will be the possible reduction in the required time, if we automate those tasks. 

\subsubsection*{Bottleneck analysis with task mining}
Bottleneck analysis is a crucial aspect of process improvement, aiming to identify and address bottlenecks that hinder workflow efficiency. Task mining, a technique that leverages process execution data from digital traces, offers a powerful approach to conduct bottleneck analysis. By capturing and analyzing user interactions with digital systems, task mining provides insights into the actual execution of tasks and reveals potential bottlenecks in the process flow.

By utilizing the provided dataset and the its recordings, task mining algorithms can identify patterns and dependencies to uncover potential bottlenecks in the process flow for specific tasks.
The dataset's rich information, including step details and contextual data, enables a holistic analysis of bottlenecks.

\section{Additional Information}
Link to the Github: \url{ https://github.com/microsoft/50BusinessAssignmentsLog }

Link to the Readme file within this repository, containing information abotu the dataset: \url{https://github.com/microsoft/50BusinessAssignmentsLog/blob/main/data/data_format.md} 

Link to the document including instruction on how to download it: \url{https://github.com/microsoft/50BusinessAssignmentsLog/blob/main/README.md}

Link to the dataset: 
\url{https://github.com/microsoft/50BusinessAssignmentsLog/blob/main/data/Recorded_Business_Tasks.csv}

Link to the license: \url{https://github.com/microsoft/50BusinessAssignmentsLog/blob/main/LICENSE}

\bibliography{sample}

\end{document}